# Compact Three-Dimensional Quantitative Phase Contrast Imager

**Vijayakumar Anand** [1,2], **Tomas Katkus** [1], **Denver Linklater**[3], **Elena P Ivanova**[3] and **Saulius Juodkazis** [1,4]

1. Center for Micro-Photonics and ARC Training Centre in Surface Engineering for Advanced Materials (SEAM), Faculty of Science, Engineering and Technology, Swinburne University of Technology, Hawthorn, VIC 3122, Australia;
2. Melbourne Centre for Nanofabrication, ANFF, 151 Wellington Road, Clayton, VIC 3168, Australia.
3. Department of Physics, RMIT, GPO Box 2476, Melbourne VIC 3001, Australia.
4. Tokyo Tech World Research Hub Initiative (WRHI), School of Materials and Chemical Technology, Tokyo Institute of Technology, 2-12-1, Ookayama, Meguro-ku, Tokyo 152-8550, Japan.
* Correspondence: sjuodkazis@swin.edu.au; Tel.: +61-39214 8718.

**Abstract:** Detection of phase variations across optically transparent samples is often a difficult task. We propose and demonstrate a compact, lightweight and low cost quantitative phase contrast imager. Light diffracted from a pinhole is incident on a thick object and the modulated light is collected by an image sensor and the intensity pattern is recorded. Two optical configurations namely lens-based and lensless cases are compared. A modified phase-retrieval algorithm is implemented to extract the phase information of the sample at different axial planes from a single camera shot.

**Keywords:** Quantitative phase contrast imaging; phase retrieval; three-dimensional imaging; lensless imaging; micro phase elements

## 1. Introduction

Phase contrast imaging techniques enable seeing thickness variations in transparent objects and have been widely used to see unstained biological samples which modulate only the phase and not the amplitude of light [1-4]. The first version of phase contrast microscope invented by Frits Zernike in 1933 consisted of an annular illumination, phase and amplitude modulating ring elements in sequence to reduce the background light and highlight the phase profile with an improved contrast [1]. Later, a modified version of the phase contrast microscope called a differential interference contrast microscopy was developed by Georges Nomarski nearly two decades later [2]. In this technique, two orthogonally polarized, mutually coherent and spatially separated light waves are interfered and the relative phase differences are imaged. In subsequent studies, interferometry and digital holography techniques evolved into reliable phase imaging methods [3,4]. In most of the interference based phase imaging applications, a coherent light source is split into two: one of the beams passes through the sample while the other beam is used as a reference. The resulting inteferogram is processed to extract the phase profile of the object. In all the above techniques [1-4], many optical components such as lenses, prisms, beam splitters, polarizers, etc., are needed for beam splitting, combining and for creation of interference patterns.

In 2004, a phase contrast imaging technique was introduced [5] in which a collimated laser light was modulated by a phase sample and the corresponding intensity pattern was recorded. The well-known Gerchberg-Saxton (GS) algorithm was implemented to retrieve the phase of the object. This is one of the most economical versions available for phase contrast imaging applications. Later, multiple wavelengths were used to improve the convergence of the phase retrieval algorithm [6]. In all the above studies, phase imaging in only a single plane was demonstrated. In this manuscript, we propose and demonstrate a three dimensional quantitative phase contrast imaging technique using a partially coherent light in two configurations: with lens and lensless, suitable for biomedical imaging and imaging of micro optical elements.

## 2. Methodology

The optical configurations with and without lens for phase contrast imaging are shown in Fig. 1(a) and 1(b) respectively. The illumination system consists of a light emitting diode and a focusing lens $L_1$ with a focal length $f_1$ which illuminates a pinhole critically. In Fig. 1(a), the light diffracted from the pinhole is collimated using a refractive lens $L_2$ with a focal length of $f_2$ located at a distance of $f_2$ from the pinhole. The collimated light is incident on a transparent sample located at a distance of $d_1$ from the lens $L_2$. The intensity of the modulated light is recorded using an image sensor located at a distance of $d_2$. The complex amplitude before the lens $L_2$ is given as $C_1\sqrt{I_o}S(f_2)$, where $C_1$ is a complex constant, $\sqrt{I_o}$ is the amplitude of light emitted from the point object and $S(f_2)$ is the spherical phase factor given as $S(f_2) = \exp[j2\pi R(f_2)/\lambda]$, where $R(f_2) = (x^2 + y^2 + f_2^2)^{1/2}$. The phase of the lens $L_2$ is given as $\exp[-j2\pi R(f_2)/\lambda]$. The complex amplitude after the lens $L_2$ is $C_2\sqrt{I_o}$, where $C_2$ is a complex constant. The transparent sample has a phase profile given as $\Phi_s$ and so the complex amplitude after the sample is $C_3\sqrt{I_o}\exp(-j\Phi_s)$ where $C_3$ is a complex constant. The complex amplitude at the image sensor is given by $C_3\sqrt{I_o}\exp(-j\Phi_s)\otimes S(d_2)$, where, '$\otimes$' is a 2D convolutional operator. The intensity recorded by the image sensor can be given as $|C_3\sqrt{I_o}\exp(-j\Phi_s)\otimes S(d_2)|^2$.

In Fig. 1(b), the case is slightly different without the lens. The light diffracted from the pinhole is incident on the sample located at a distance of $d_1$ from it. The light modulated by the sample is recorded by an image sensor located at a distance of $d_2$ from the sample. The complex amplitude incident on the sample is given as $C_1\sqrt{I_o}S(d_1)$ and after the sample is $C_4\sqrt{I_o}S(d_1)\exp(-j\Phi_s)$. The complex amplitude at the image sensor is given by $C_4\sqrt{I_o}S(d_1)\exp(-j\Phi_s)\otimes S(d_2)$. The intensity recorded by the image sensor can be given as $|C_4\sqrt{I_o}S(d_1)\exp(-j\Phi_s)\otimes S(d_2)|^2$. In both configurations, it is not possible to know the phase profile of the sample directly. The phase of the sample is calculated using an iterative GS algorithm [8]. In both configurations, there are two planes of interest: sample plane and sensor plane and in both cases, it is assumed that the sample introduces only phase modulation and no amplitude modulation. Therefore, in the GS algorithm, the initial input amplitude and phase profile is constant for both cases. If the values of $d_1$ and $d_2$ are known, then the initial input phase can be synthesized. For larger distances, a quadratic phase approximation of the spherical phase is sufficient. For universal application, the spherical phase factor has been used. The schematic of the GS algorithm is shown in Fig. 2. In the lensless case, it is necessary to increase the distance $d_2$ in the GS algorithm to compensate the spherical phase factor present in the sample plane [7].

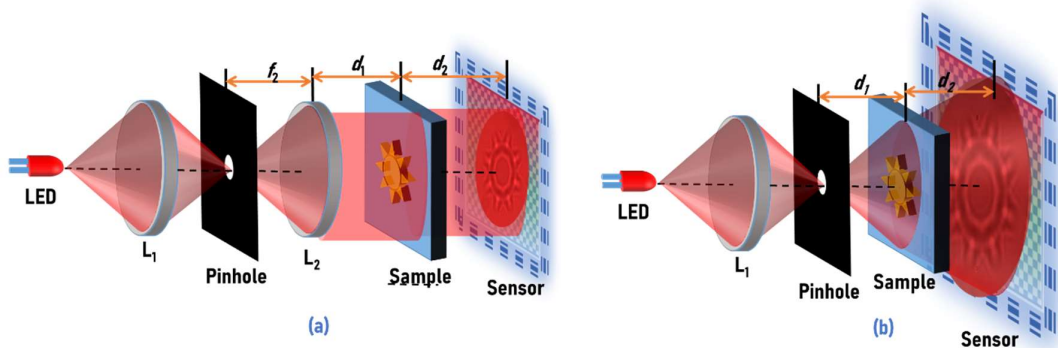

**Figure. 1** Optical configuration of the phase imager (a) with lens and (b) without a lens.

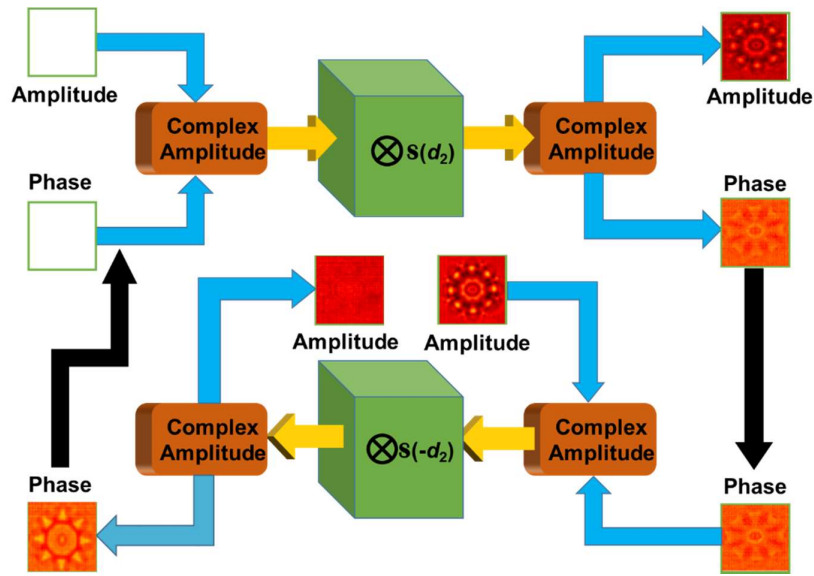

**Figure. 2** Modified Gerchberg Saxton algorithm

Calculations were carried using MATLAB on a computer (Intel Core i5-8250U CPU @ 1.60GHz and 1.80 GHz). The initial guess of the amplitude and phase inputs of the algorithm are constants and after convolution with the spherical phase factor, the phase is preserved while the amplitude was replaced by the square root of the recorded intensity pattern. The complex amplitude with the experimental amplitude profile is convolved with the negative of the spherical phase factor and the output phase is preserved while a constant amplitude is added for the next iteration.

## 3. Experiments and Results

An experimental set up based on Fig. 1(a) and (b) is built using a LED ($\lambda_c$ = 530 nm, FWHM = 33 nm) and a pinhole with a diameter of 100 μm. In the first optical configuration, the distances $d_1 \approx d_2 \approx$ 3 cm. The experiment was first carried out using a single thin sample. The optical microscope images of the phase objects - Swinburne logo and two stars and the word 'Nan' fabricated using electron beam lithography (EBL; Raith 150²) [9, 10] in PMMA 950K resist on Indium Tin Oxide (ITO) glass substrates with 1.1 mm thickness are shown in Figs. 3(a)-3(c), respectively. The thickness $t$ of PMMA resist was about 800 nm and its refractive index is closely matching that of glass. The developed patterns after EBL exposure (Fig. 3) represent mostly phase structures (a jump in PMMA thickness) which were used for the optical imaging experiments. Thick objects were built by attaching two different thin phase objects with a 1 cm spacer as shown in Fig. 3(d).

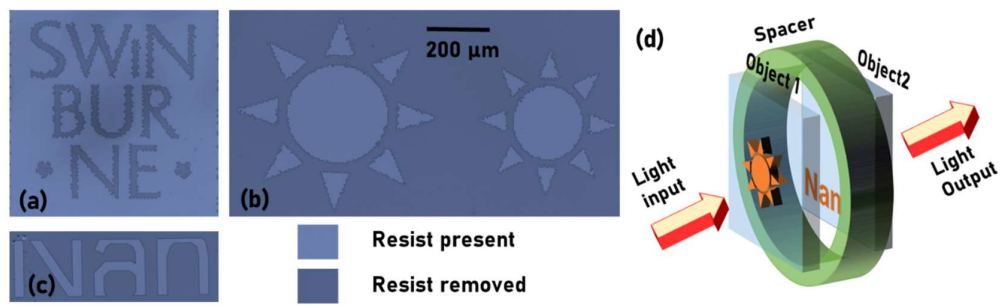

**Figure. 3** Optical microscope images of (a) Swinburne logo, (b) two stars and (c) object 'Nan' fabricated using electron beam lithography. The darker regions indicate resist removed and the brighter regions indicate resist remaining. (d) The configuration for building thick objects from two thin objects.

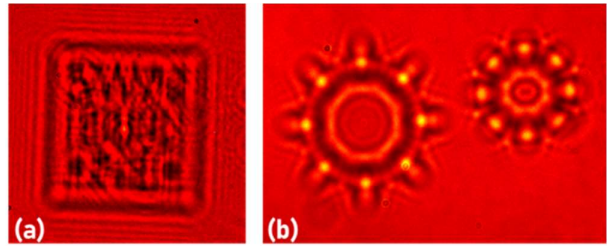

**Figure. 4** Intensity patterns recorded by the image sensor for (a) Swinburne logo with a lensless system and (b) stars with lens based system.

The images of the intensity patterns for the Swinburne logo and two stars recorded by the image sensor (Thorlabs DCU223M, 1024 x 768 pixels, pixel size = 4.65 µm) for the lensless system and lens based system respectively are shown in Figs 4(a) and 4(b), respectively. The presence of the glass substrate and the location of the CCD chip inside the image sensor may result in a small error in the distance measurement. Therefore, a test run is executed with a known object in the GS algorithm by varying the distance $d_2$ over a set of values, until an optimal reconstruction is obtained. Once the optimal reconstruction is obtained, the same value of $d_2$ can be used for the reconstruction of other objects mounted in the same axial distance. The reconstructed images of the second star and Swinburne logo and the corresponding phase profiles are shown in Fig. 5. The dip and protrusion in the Swinburne logo and star are visible respectively.

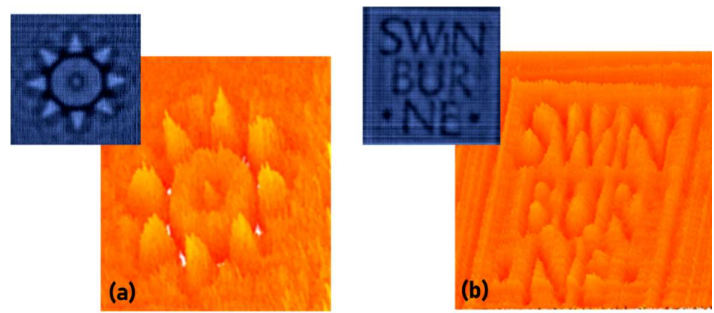

**Figure. 5** Reconstructed images and their phase profiles (top left insets) for (a) star and (b) Swinburne logo.

Three dimensional phase image reconstruction is discussed next. A thick phase object is created using two objects namely the star and the object 'Nan' separated by a distance of 1 cm. The distance $d_2$ was varied by 1 cm from the initial value in the phase retrieval algorithm and the results of reconstruction using the system without lens when the star and the object 'Nan' are in focus are shown in Fig. 6(a) and 6(b) respectively.

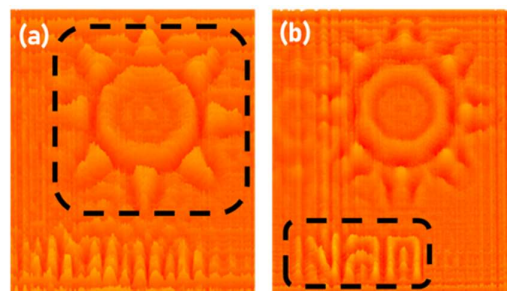

**Figure. 6** Reconstruction results when the location of (a) star and (b) object 'Nan' was digitally refocused. The focused object is indicated by a box.

**4. Discussion**

The phase difference step of imaged objects (in air) was given as $2\pi t(n-1)/\lambda$ which is approximately $1.5\pi$ for PMMA of refractive index 1.5 and thickness 800 nm at 530 nm wavelength. This is a small phase change typical for bio-imaging microscopy carried out in solution (refractive index 1.33 of water) and for micro optical elements. The lateral resolution of imaging is limited by the pixel size of the camera rather than the NA of the lens as in the case of regular microscopy. The field of view is limited by the physical size of the image sensor. There is some degree of freedom in the lensless case with a diverging wave illumination of the sample, where the distance $d_2$ can be adjusted to trade-off resolution of imaging with the field of view. The phase retrieval algorithm required as low as only 4 iterations to converge and the execution time was only 2.4 seconds. It must be noted that the limitation will not apply for real time observation of a biological event or fabrication of a micro optical element unless the event needs a reactive real time intervention.

**5. Summary, Conclusion, Outlook**

A compact, single camera shot, three dimensional phase imaging technique using a partially coherent light source has been proposed and demonstrated. The resolution of imaging was approximately 10 μm. The number of iterations required for the reconstruction is only 4, which makes the system temporally faster, suitable for recording live events once the reconstruction distance has been determined during the preliminary training. The proposed technique has many advantages in comparison to regular phase contrast microscopes. The proposed method does not require many optical components but only a pinhole and a web camera and works with an LED which makes the system low-cost by at least an order of magnitude in comparison to conventional microscopes. Secondly, in a conventional microscope, only a particular plane can be focused at a time but in the proposed system a single recorded image can be digitally focused and refocused to any plane of interest. Consequently, the proposed method can be used to monitor multiple planes of an object simultaneously, real time.

This method is promising for bio-microscopy of cells and particularly for monitoring the interaction of bio-membrane with the nanotextured surface to reveal mechano-bactericidal action [11]. Modifications inside transparent materials, phase changes, refractive index alterations during femtosecond laser writing of micro-optical elements [10] can be augmented with in situ monitoring using the proposed method. These are directions that will be explored in the application of this phase contrast imager.

**Author Contributions:** Conceptualization, V.A; Fabrication, V.A and T.K.; experiments, V.A and D.L.; validation, S.J and E.P.I; resources, S. J; writing—original draft preparation, V.A.; writing—review and editing, S. J; supervision, S. J.; project administration, S. J.; funding acquisition, S. J. All authors have read and agreed to the published version of the manuscript.

**Funding:** NATO grant No. SPS-985048 and the Australian Research Council Discovery grant DP190103284 are acknowledged for funding. This work was performed in part at the Swinburne's Nanofabrication Facility (Nanolab).

**Conflicts of Interest:** The authors declare no conflict of interest